# Diseño de un espectrofotómetro UV-VIS de bajo costo para la industria bioquímica: una revisión

## Design of a low-cost spectrophotometer UV-VIS for biochemistry industry: a review

L. Moreno-Villalba [a], F. J. Ávila-Camacho [b*], I. G. Montes-Cerón [c], A. Garrido-Hernández [c], C. A. Cardoso-Chávez [b], C. J. Pacheco-Piña [b]

[a] *División de Ingeniería en Informática, Tecnológico de Estudios Superiores de Ecatepec, 55210, Ecatepec, Estado de México, México.*
[b] *División de Ingeniería en Sistemas Computacionales, Tecnológico de Estudios Superiores de Ecatepec, 55210, Ecatepec, Estado de México, México.*
[c] *División Químico Biológica, Universidad Tecnológica de Tecamac, 55740, Tecamac, Estado de México, México.*

**Resumen**

Los sistemas de espectrofotometría requeridos comúnmente tanto en áreas de investigación como en industrias químicas y de materiales resultan ser equipos costosos para laboratorio, muchas ocasiones se vuelve inviable la adquisición de estos; además, en varias de las aplicaciones también se requiere portabilidad para realizar mediciones en campo y que los mismos pudieran ser de amplio espectro. En este trabajo se llevó a cabo una revisión de diversos proyectos de desarrollo de dispositivos de espectrofotometría que van desde prototipos académicos hasta productos industriales utilizando distintas estrategias de implementación con el fin de conocer la factibilidad de crear dispositivos portátiles y de amplio espectro o productos específicos para un sector en particular, pero con la intención de lograr una alta precisión y funcionalidad. La investigación permitió concluir con una aportación novedosa para el desarrollo de prototipos de espectrofotometría de gran alcance para industrias o sectores específicos y de bajo costo.

*Palabras Clave:*
Espectrofotómetro, Técnicas de medición, Medición de adsorción, Luz ultravioleta visible.

**Abstract**

The spectrophotometry systems commonly required both in research areas and in the chemical and materials industries turn out to be expensive laboratory equipment, in many cases the acquisition of these becomes unfeasible; besides, in several applications portability is also required to perform measurements in the field and that they could be broad spectrum devices. In this work, a review of various development projects for spectrophotometry devices was carried out, ranging from academic prototypes to industrial products using different implementation strategies in order to know the feasibility of creating portable and broad-spectrum devices or specific products for a particular sector, but with the intention of achieving high precision and functionality. The research allowed to conclude with a novel contribution for the development of powerful spectrophotometry prototypes for specific, low-cost industries or sectors.

## 1. Introducción

Los sistemas de espectrofotometría se utilizan en una gran variedad de industrias, así como en campos de investigación relacionados con la química, bioquímica, química molecular y en general con las ciencias de los materiales para diversas aplicaciones como en la medición de la concentración de gases, los niveles de adsorción y absorción en compuestos y materiales (Barth, 2000), (Berendzen & Braunstein, 1990), (Caswell & Spiro, 1987), (Vernon, 1960).

El análisis ambiental y de salud ha ido en aumento en la última década debido a los efectos destructivos que se están ocasionando en el medio ambiente, provocando que sea necesario el análisis de sustancias dañinas tanto para el ser humando como para el entorno. Este proceso puede ser muy deficiente debido a que es necesario recolectar muestras de un sitio para luego ser transportadas a un laboratorio y así poder ser analizadas. Los métodos que se utilizan para analizar las muestras pueden ser muy costosos o muy lentos e inclusive pueden no ser exactos ya que la muestra puede contaminarse en el transporte (Ho, Robinson, Miller, & Davis, 2005), igualmente los espectrofotómetros convencionales pueden llegar a ser muy costosos, con precios que superan los $6000 dólares (PROLAB (Proveedor de Laboratorios S.A. de C.V.), 2021). Para evitar los inconvenientes descritos, a lo largo del tiempo, se ha tratado de miniaturizar el espectrofotómetro para poder hacerlo más práctico y versátil, además de económico (Marle & Greenway, 2005).

Debido a que varias de las aplicaciones mencionadas anteriormente, como en el caso del análisis de nanopartículas metálicas crecidas en líquidos (Sebastián & Pablo, 2015), se requieren instrumentos de alta precisión y confiabilidad, es necesario importar equipo a costos elevados. Estos equipos normalmente son grandes y voluminosos por lo que se requiere contar con espacios adecuados para su operación. En algunas ocasiones, para aplicaciones de campo, se vuelve necesario contar con equipo portable para realizar mediciones en sitio y para ello es importante contar con equipos portátiles de amplio espectro (Asher, Ludwig, & Johnson, 1986), (Hodgkinson, Shan, & Pride, 2006), (Lackner, 2007)

Existen varios proyectos tanto comerciales como académicos para construir espectroscopios y espectrofotómetros portables y de bajo costo (Chng & Patuwo, 2020; Kovarik, Clapis, & Romano-Pringle, 2020; Place, 2019; Sölvason & Foley, 2015), aunque en la mayoría de los casos son proyectos educativos, no presentan mucho detalle en su diseño y construcción por lo que no permite analizar su aplicación y nivel de precisión.

Un ejemplo en los valores de precisión deseables para un espectrofotómetro se puede observar en las tablas de referencia utilizadas para calibrar instrumentos como el UV-VIS modelo INM/GTM-FR-E/01. Una de ellas se muestra en la tabla 1.

Tabla 1: Ejemplo de presentación de resultados para la escala de longitud de onda

| ʎ (nm) | Valor medido (nm) | Valor MRC (nm) | Error (nm) |
|---|---|---|---|
| 1 | 278.94 | 279.28 | -0.34 |
| 2 | 287.23 | 287.58 | -0.35 |
| 3 | 333.30 | 333.92 | -0.62 |
| 4 | 347.41 | 347.99 | -0.58 |

Para validar los resultados de la calibración, se deberá especificar, por cada longitud de onda, el valor promedio de las mediciones, el valor del material de referencia certificado (MRC), el error y la incertidumbre asociada. Los valores que se presentan en la tabla 1 se deberá realizar para la transmitancia y la absorbancia.

En la parte comercial existen desarrollos portables que utilizan una técnica reflectiva para obtener información cualitativa de las muestras y en algunos casos utilizando el teléfono celular (Das, Wahi, Kothari, & Raskar, 2016; SCiO Product, 2020).

En (Gallegos et al., 2013) se utiliza la cámara de un celular como espectrómetro el cual incluye también un biosensor de cristal fotónico, para ello requiere de un tripee para fijar el teléfono y alinearlo con los componentes ópticos para asegurar mediciones precisas y repetibles. La luz originada que es de amplio espectro se polariza de forma lineal para pasar por el biosensor, el cual refleja una banda de longitudes de onda reducida, al mismo tiempo, una rejilla de difracción distribuye las longitudes de onda restantes dentro de la cámara para obtener un espectro de transmisión de alta resolución. Adicionalmente, en ese trabajo se creó una aplicación móvil para convertir las imágenes que detecta la cámara en el espectro de transmisión que se requiere.

El analizar la reflexión de la luz tiene algunas ventajas como el hecho de poder medir sólidos y no requerir de una preparación previa de las muestras, pero no podría generar información cuantitativa que es necesaria para obtener valores absolutos, dado que el espectrómetro no recibiría toda la luz reflejada (Varra et al., 2020).

En este trabajo se lleva a cabo una revisión sobre las diversas opciones de diseño para un espectrofotómetro con el fin de realizar un análisis para la implementación de un dispositivo portable de bajo costo que además sea ajustable y de rápida respuesta para un espectro completo de longitud de onda.

## 2. Antecedentes

Un espectrofotómetro es un dispositivo que se utiliza para el análisis cuantitativo de una sustancia química, biológica, orgánica o inorgánica. Para ello, se mide la cantidad de luz que es absorbida por el compuesto presente en una solución (Thomas & Burgess, 2017).

La región del espectro de luz clasificada como ultravioleta (UV), va de los 190 nm a los 380 nm, la luz visible (Vis), va de los 380 nm a los 750 nm y la luz infrarroja cercana (NIR por sus siglas en inglés), va de los 800 nm a los 2500 nm (Cazes, 2004), (Grinter & Threlfall, 1992), (P. Worsfold, Poole, Townshend, & Miro, 2019).

Como se describió en la introducción, existen desarrollos académicos de diseño de espectrofotómetros en los cuales su funcionamiento se centra en el espectro de luz UV y NIR, que utilizan matrices de diodos y dispositivos de par cargado (Gonzaga & Pasquini, 2010), (Noui et al., 2002). Los espectrofotómetros de rango de luz visible de bajo costo utilizan como fuente de luz principal los diodos LED, ya sea de color blanco que utiliza un monocromador para separar la luz, o con diodos LEDs (Diodo Emisor de Luz) de colores para obtener un rango de las diferentes frecuencias de luz (O'Toole & Diamond, 2008).

Dentro de las principales aplicaciones de un espectrofotómetro se tiene el uso del dispositivo para medir contaminantes en el agua (Sarkar, 1999), así como para obtener concentraciones de algunas células humanas (Bishop, Fody, & Schoeff, 2018) y para el análisis de diversas sustancias utilizadas en diversas industrias (Tzanavaras & Themelis, 2007).

Las fuentes de luz led y los fotodetectores se utilizan hoy en día para crear diferentes prototipos de dispositivos espectrofotómetros de bajo costo, compactos y versátiles que permiten obtener resultados muy similares a los espectrofotómetros convencionales (Sequeira et al., 2002).

## 3. Elementos básicos de un espectrofotómetro

### 3.1. Fuentes de luz

El avance que se ha tenido en las tecnologías para la fabricación de LED's (Diodo Emisor de Luz) proporciona ventajas para el desarrollo de espectrofotómetros compactos ya que pueden sustituir gran cantidad de otras fuentes de luz optoelectrónicas como lo son las lámparas de deuterio y tungsteno, teniendo un costo más bajo, una vida útil mayor, menor consumo de energía y se les puede encontrar en rangos de luz que

van desde la luz UV hasta la luz infrarroja (247 nm-1550 nm) (Gardolinski, David, & Worsfold, 2002), (Kuo, Kuyper, Allen, Fiorini, & Chiu, 2004), (Roithner Laser Technik GmbH, 2020), (Schubert & Kim, 2005), (Taniyasu, Kasu, & Makimoto, 2006).

Para obtener fuentes de luz confiables, es necesario que muestren estabilidad, direccionalidad, distribución de energía espectral continua y larga vida. Las fuentes que se utilizan en los espectrofotómetros son principalmente lámparas de tungsteno de bajo voltaje, lámpara tungsteno-halógena o lámpara de pulso/arco de xenón cuya utilidad se centra en la región del espectro visible al infrarrojo. 320 – 2500 nm (Jhilmer, Iris, Isabel, Camila, & Viviana, 2019).

### 3.2. Monocromador

La función principal de un monocromador es la de separar el espectro de luz y para ello utiliza una ranura de entrada y un elemento dispersor que puede ser una rejilla de difracción o un prisma, por lo que también podría requerir de lentes y/o espejos para dirigir el espectro de luz seleccionado a una ranura de salida. De acuerdo con la norma ASTM E958, el ancho de banda espectral es el intervalo de longitudes de onda de radiación en la salida del monocromador medido en la mitad de un pico del flujo de radiación detectada y con ello se determina el rasgo espectral más corto que puede detectar el espectrofotómetro (Jhilmer et al., 2019).

### 3.3. Sensores

Los sensores son transductores que convierten la energía de radiación luminosa en señal eléctrica. Es decir, la capacidad de producir una señal eléctrica cuando el sensor es bombardeado con fotones. Una de las principales características que debe cumplir es el tiempo de respuesta. (L.C. Passos & M.F.S. Saraiva, 2019). En (L.C. Passos y M.F.S. Saravia) se utilizó el sensor de luz ambiental TEMT6000 (Vishay Intertechnologies, PA, EE. UU.) (HETPRO, 2021). El dispositivo consta de un fototransistor positivo negativo (NPN) montado en una placa de circuito impreso (PCB) de 1 cm × 1 cm (Illustrationprize, 2021). Los detectores se encuentran ubicados de forma que la luz pueda ser absorbida después de traspasar la muestra y las sustancias de referencia almacenadas en el cuarzo.

### 3.4. Algoritmos de control y microcontroladores

En la mayoría de los proyectos analizados se utilizaron microcontroladores Atmega 138p con la interfaz de Arduino para realizar la programación de los algoritmos. Como primer paso se realiza una rutina de inicialización, consiste en encontrar una posición inicial del monocromador utilizando una referencia física. Utilizando micro pasos el motor recorre todas las longitudes de onda de todo el espectro visible. Para cada paso que da el motor se almacena el voltaje que se recibe luego de procesar la luz que pasa a través de la muestra. El voltaje es leído por el convertidor analógico digital del microcontrolador. Por último, el valor que se obtiene, valor de absorbancia, se calcula y almacena en una tabla.

### 3.5. Configuración de los elementos

En los principales desarrollos encontrados, los componentes se ensamblaron en una caja negra fabricada mediante impresión 3D con dimensiones de 15 cm x 22 cm x 8 cm y un espesor de pared y tapa de 4 mm. Para tener una buena dispersión de la luz, la rejilla de difracción se ubica a 18cm de la cámara de muestra. Los detectores se ubican en un espacio con aislamiento lumínico para disminuir la interferencia de luz externa en los análisis (D. González-Morales et al., 2020a).

La figura 1 muestra el ensamblado de un espectrofotómetro construido por (D. F. González-Morales, López-Santos, & García-Beltrán, 2018) y donde se realizó una prueba inicial para medir la concentración del sensor con cumarina. La segunda prueba consistió en medir la concentración de mercurio a diferentes concentraciones. Las pruebas fueron realizadas en paralelo con el espectrofotómetro comercial Cary 60 (Agilent Technologies, Santa Clara, CA, EE. UU.), (D. González-Morales et al., 2020a).

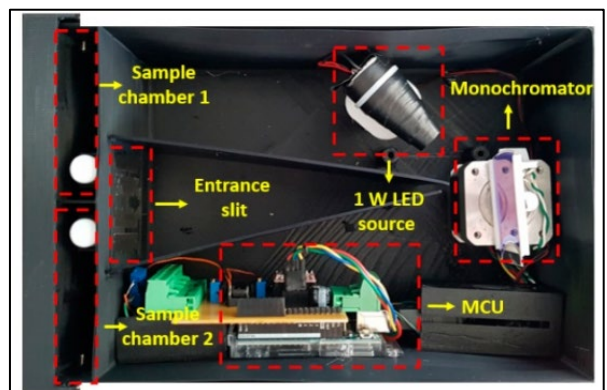

**Figura 1. Ensamblado de un espectrofotómetro. Imagen tomada de (D. González-Morales et al., 2020b)**

Como resultado obtuvieron las ligeras diferencias entre los resultados del dispositivo que desarrollaron y el equipo comercial Cary 60, por lo que el espectrofotómetro desarrollado fue realmente funcional, practico y económico, cumpliendo los requerimientos propuestos por sus desarrolladores.

### 4. Alternativas a los espectrofotómetros comerciales

### 4.1. Espectrofotómetro de bajo costo

El modelo que se describe es una contribución al hardware de código abierto que se encuentra continuamente creciendo (Baden et al., 2015; Maia Chagas, 2018). Los planos están disponibles en GitHub (Vasco Ribeiro Pereira, 2019).

Para su construcción se utilizaron diodos LED con diferentes longitudes de onda. El costo de cada LED fue de menos de 0.10 € por cada uno.

Los autores se encargaron de desarrollar dos prototipos, uno realizado con una impresora 3D y otro utilizando bloques de lego, a continuación, se explica la construcción de cada uno.

La muestra se coloca en portaobjetos creado en forma de cubeta y con una sección intercambiable que sostiene al LED y que se alimenta con una pila CR2032, como se muestra en la figura 2. Ambas partes se encuentran impresas en plástico 3D. La pieza del soporte de objetos tipo cubeta tiene un agujero y una pieza de goma que le permite colocar la muestra sin que esta se mueva y evita que la luz exterior llegue al sensor. Para la versión de lego la apertura frontal se alinea con la apertura trasera y, cuando está instalada, con el teléfono inteligente muestra el mecanismo del interruptor utilizado en la parte intercambiable. Cuando la batería está desconectada, el LED se apaga y cuando la batería toca las patas del LED este se enciende (Pereira & Hosker, 2019).

La aplicación que se utilizó para poder calcular la absorbancia de acuerdo a las mediciones de la luz se utilizó la aplicación

"Shoebox spectrophotometer" que se encuentra disponible en Google Play sin costo alguno (Hosker, 2018). La aplicación permite medir la luz en tiempo real por lo que permite el posicionamiento correcto del portaobjetos, con el LED alineado al sensor de luz del celular (Ver figura 2). Para poder saber que el sistema se encuentra bien protegido de la luz externa, al apagar el LED, la medición en tiempo real debe de marcar un cero (Pereira & Hosker, 2019).

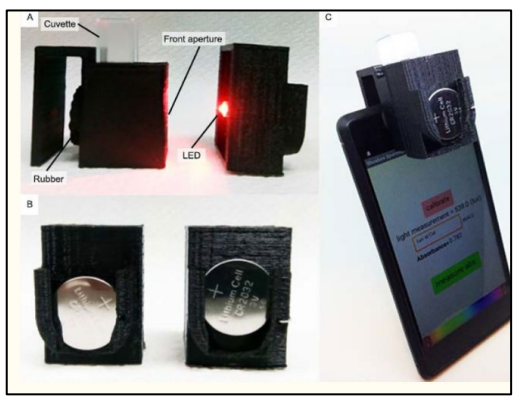

**Figura 2. Impresión en 3D del espectrofotómetro económico. Imagen tomada de (Pereira & Hosker, 2019)**

La figura 3 muestra el espectro realizado con piezas de Lego con la finalidad de evaluar la funcionalidad del espectrofotómetro, se probó la absorbancia en tres diferentes soluciones violeta cristal a 587nm, naranja de metilo a 465nm y p-nitroanilina a 400nm.

Se compararon los resultados obtenidos por el espectrofotómetro desarrollado junto con tres diferentes espectrofotómetros comerciales. Se observó una desviación negativa de la ley de Beer en todos los espectrofotómetros comerciales a concentraciones de p-nitroanilina por encima de 200 µM, igualmente el espectrofotómetro del teléfono se encontró una desviación similar (Pereira & Hosker, 2019).

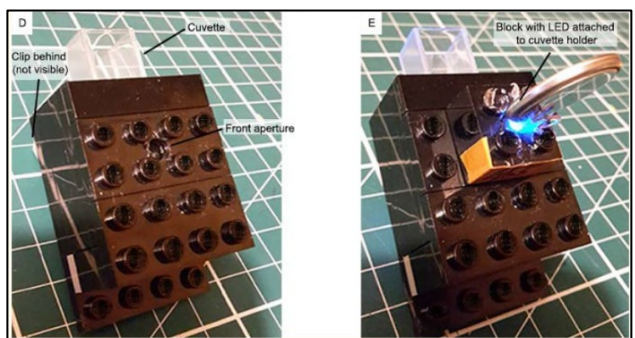

**Figura 3. Espectrofotómetro realizado con piezas lego. Imagen tomada de (Pereira & Hosker, 2019)**

Como se describe, los resultados fueron satisfactorios para un espectrofotómetro realmente económico y que se puede realizar sin conocimientos de programación o configuración de hardware, por lo cual es una buena opción de elaboración para personas que se enfocan en el campo de la química y necesitan de este instrumento.

### 4.2. Fotómetro UV simple y económico

En este proyecto (Apostolidis, Klimant, Andrzejewski, & Wolfbeis, 2004) utilizaron un fotómetro que utiliza diodos LED UV para la fuente de luz y a su vez también como detector, para mediar la intensidad UV. Utilizaron un LED de 355nm para la fuente de luz, cuyo precio ronda en los $30 dólares y también representa poco riesgo a la salud dada la longitud de onda.

El fotómetro que desarrollaron considera que los LED pueden ser utilizados como detectores de la luz cuando las longitudes de onda son iguales o más cortas de las que emiten (Lau et al., 2004; Macka, Piasecki, & Dasgupta, 2014; W, 1973). El fotómetro es simple y todas las partes pueden moverse ya que se encuentra construido con ladrillos de lego.

El fotómetro que se desarrolló se puede ensamblar rápidamente y solo necesita una pequeña cantidad de circuitos que son la batería de 4.5 V, una resistencia, cables y los dos LED. La resistencia permite el funcionamiento óptimo del rango de luz del LED. Los ladrillos legos son muy prácticos ya que permiten la correcta alineación de los LED y debido a esas características se han construido otros modelos con ellos (Albert, Todt, & Davis, 2012; Campbell, Miller, Bannon, & Obermaier, 2011). El autor también menciona que puede utilizarse otro tipo de estructuras como una impresa en 3D (Anzalone, Glover, & Pearce, 2013).

El fotómetro funciona con un LED como detector sin circuitos. La medición de voltaje, se realiza directamente a través del LED y el multímetro.

La parte superior de un ladrillo Lego (2 × 2) está cortado para acomodar una cubeta. Cada LED se coloca en un ladrillo Lego con agujero (2 × 1 con agujero). Una fuente de alimentación de 4,5 V (3 × Baterías de 1,5 V) cableadas en serie con una resistencia adecuada para el LED en particular (normalmente en el rango de 10 - 50 Ω para un UV-LED de 355 nm) alimenta el fotómetro. La intensidad de la emisión de un LED depende del voltaje aplicado y la temperatura del LED (Kvittingen, Kvittingen, Sjursnes, & Verley, 2016).

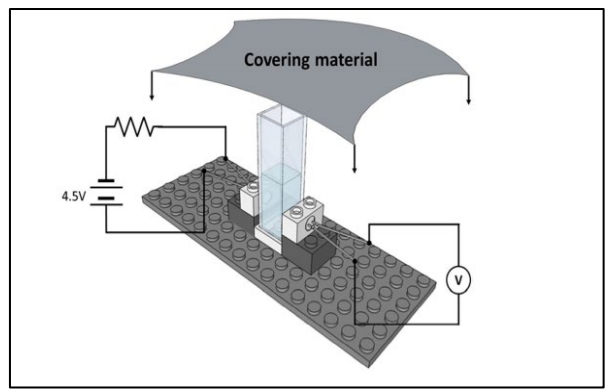

**Figura 4. El fotómetro funciona con un LED como detector sin circuitos. La medición, voltaje, se realiza directamente a través del LED. Imagen extraída de (Kvittingen et al., 2016)**

La figura 4 ilustra la adaptación de un cargador de celular con la finalidad de tener una fuente de alimentación estable, ya que se necesitan de 15 a 30 minutos para que el LED funcione en condiciones estables, pero con el tiempo deteriora el estado de la batería lo cual no permite el funcionamiento óptimo del LED.

Las mediciones de la trasmitancia se realizan con ayuda de un multímetro digital con una resistencia de al menos 10 MΩ que se conecta directamente al LED receptor. Los voltajes comúnmente varían entre 1 a 100 mV. De ser necesario se puede colocar algún material negro como cubierta para evitar la luz externa que pueda alterar los resultados (Kvittingen et al., 2016).

Se prepararon soluciones estándar de cinamaldehído en el rango de 50 - 1500 mg / L en etanol / agua (50/50%). El cinamaldehído (1 - 2%) se aisló mediante destilación al vapor en una columna Hickmann de canela en polvo y corteza de canela (aproximadamente 500 mg) y se diluyó en etanol al 50%.(Kvittingen et al., 2016). Se comparó la sustancia con el fotómetro UV que desarrollaron y un espectrofotómetro comercial Hitachi U-2000 (Biocompare, 2002).

La figura 5 muestra que la absorbancia del fotómetro UV es ligeramente superior a la que se obtuvo con el espectrofotómetro comercial, esto debido a que la longitud de onda varía entre 353 a 360 nm.

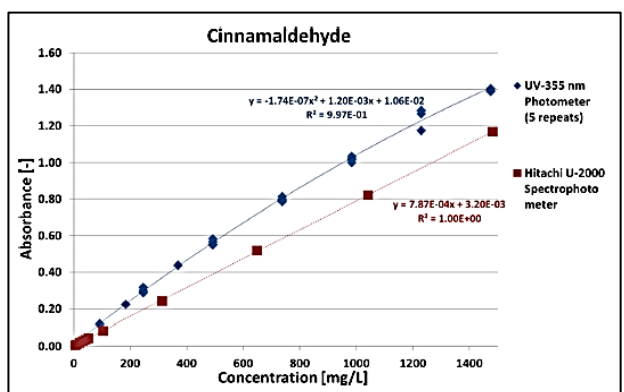

**Figura 5. Absorbancia de luz de un LED UV (355 nm) por cinamaldehído (en etanol al 50%), utilizando el fotómetro UV y un espectrofotómetro comercial Hitachi U-2000. Imagen tomada de (Kvittingen et al., 2016)**

*4.3. Mini espectrofotómetro para determinación de glucosa plasmática.*

A diferencia de los espectrofotómetros, una fuente de luz del mini espectrofotómetro se reemplaza por una luz diodo emisor (LED) y un micropocillo fabricado a base de polímero se utiliza como cubeta. A validar el prototipo de espectrofotómetro de reducción de tamaño, la eficiencia y confiabilidad para investigar las determinaciones de glucosa (Chaianantakul et al., 2018).

Generalmente se usa un espectrofotómetro para determinar la cantidad relativa de moléculas de interés midiendo la intensidad de la luz que se absorbe parcialmente. Normalmente, el espectrofotómetro está equipado con una lámpara de tungsteno o deuterio como fuente de luz.

Sin embargo, un espectrofotómetro convencional es grande y complejo, por lo que no es adecuado para análisis de campo. El dispositivo y los accesorios son caros y requieren el mantenimiento de un técnico capacitado y certificado. Desarrollo avanzado reciente en una fuente de luz de equipo

Se ha demostrado que el diodo emisor de luz (LED) se utilizó con éxito como luz fuente en lugar de sistema óptico complejo. El LED es más estable, confiable y de mayor duración-tiempo, y más pequeño que otras fuentes de luz. La fuente de luz LED se ha utilizado para determinación de creatinina colorantes alimentarios y hierro, hidrocarburos aromáticos, zinc y cobre.

El mini espectrofotómetro (W 16 cm x D 15 cm x H 6 cm) utilizado en este estudio fue desarrollado por Synchrotron Light Research Institute (Organización Pública), Tailandia.

El sistema consta de 8 fuentes de luz LED que incluyen violeta (398 nm), azul (465 nm), verde (517nm), amarillo (593 nm), naranja (604 nm), rojo (641 nm), blanco y rosa (componentes RS, Tailandia), micropocillos, sensor de luz TSL235R (Texas Instruments, EE. UU.) Y se muestra como se ilustra en la figura 6 y en la figura 7. La fuente de luz LED verde con una longitud de onda de emisión máxima de 517 nm se utilizó para la determinación de glucosa (Chaianantakul et al., 2018).

Los espectrofotómetros convencionales utilizados en el estudio comparativo fueron Shimadzu UVEspectrofotómetro 1601 (Shimadzu Corporation, Japón) y CE 1021 UV / Espectrofotómetro Visible (CECIL Instruments Limited, Reino Unido) (Chaianantakul et al., 2018).

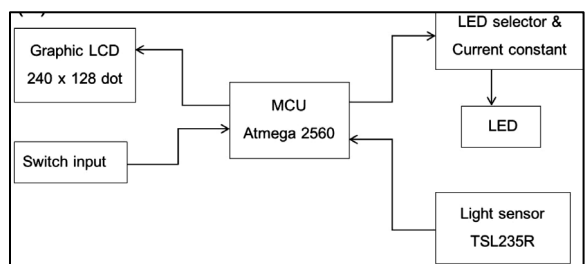

**Figura 6. Estructura del espectrofotómetro (Chaianantakul et al., 2018)**

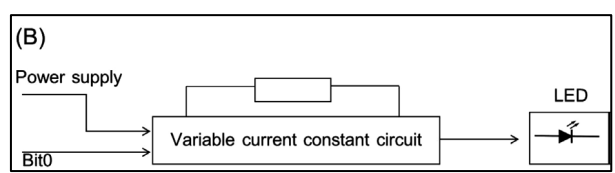

**Figura 7. Sensor de luz (Chaianantakul et al., 2018)**

*4.4. Espectrofotómetro de radiación visible tipo casero*

El espectrómetro propuesto utiliza como detector óptico una cámara web conectada a la computadora, con ello se lleva a cabo la adquisición, y registro de los espectros. Como dispersor se utiliza una porción de un DVD virgen, con lo que se obtiene una difracción y la luz se transmite al dispersor mediante un trozo de fibra óptica acrílica. Los componentes del instrumento están montados rígidamente sobre una placa de aluminio pintada de color negro (Rossi, Elías, & López, 2013).

Adicionalmente se utiliza como detector una cámara Web Logitech Quickcam Pro 9000 ®, conectada a la computadora con sistema operativo Linux ®, Ubuntu 12.04 LTS ® (Rossi et al., 2013).

El resultado es un espectrómetro robusto que no tiene partes móviles. En la figura 8 se muestra que los tres componentes están montados fijamente sobre una placa de aluminio de 240 x 100 x 50 mm, que constituye el banco óptico y que se ha ennegrecido con pintura negra no reflectora. Para fijar cada componente se ha empleado pequeñas láminas de metal, también ennegrecidas. En la figura 9 se muestra que el montaje de los tres componentes, está confinado en una carcasa cuyo interior y exterior se han cubierto también con la misma pintura (Ver figura 8 y figura 9) (Rossi et al., 2013).

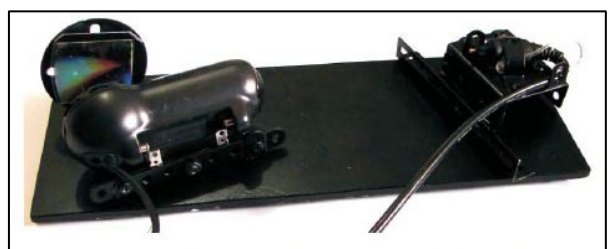

**Figura 8. Placa de aluminio conteniendo los tres componentes del espectrofotómetro (Espectrómetro Para Radiación Visible Hecho En Casa, de Bajo Costo y Altas Prestaciones, n.d.)**

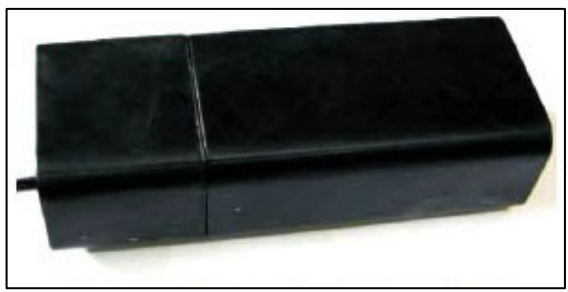

**Figura 9. Vista exterior del espectrofotómetro (Espectrómetro Para Radiación Visible Hecho En Casa, de Bajo Costo y Altas Prestaciones, n.d.)**

*4.5. Prototipo con porta celda para realizar espectrometría de absorción y fluorescencia moleculares.*

En la figura 10 se ilustra que el recipiente donde se colocan las muestras se construyó con una masa de la marca MOLDIMIX, utilizando un molde creado con una cubeta de polietileno para evitar que se adhiera a la masa. En la parte externa se utilizó un contenedor de rollo fotográfico, que también tiene forma cilíndrica, al cual, una vez que se retiraron ambos moldes, se hicieron agujeros para la entrada y salida de luz y un agujero en ángulo recto con los dos anteriores, para las medidas de fluorescencia y turbidez. (Rossi et al., 2013).

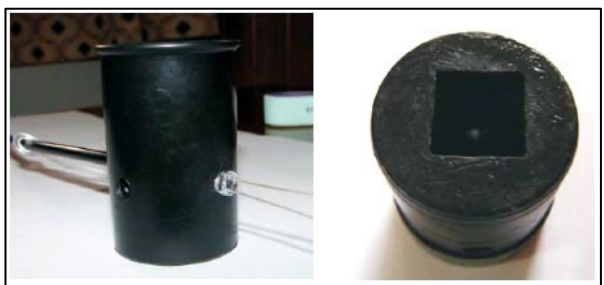

**Figura 10. Ejemplos de dispositivos porta-celda(Espectrómetro Para Radiación Visible Hecho En Casa, de Bajo Costo y Altas Prestaciones, n.d.)**

Todo el software empleado en el presente trabajo es de licencia libre. Para enfocar y configurar la cámara web, se utilizó el software QT v4l2 (Rossi et al., 2013).

Para la adquisición de datos, que consiste en la toma de un número dado de fotografías, en forma automática, se utilizó el software denominado streamer (Gerd Knorr <kraxel@bytesex.org>). El procesamiento de datos se realizó con una aplicación en forma de macro, utilizando el software ImageJ (http://imagej.nih.gov/ij/). Finalmente, para la calibración de los espectros en longitudes de onda, se empleó la hoja de cálculo de la suite Open Office (Rossi et al., 2013).

Como fuente de radiación se utilizó un diodo emisor de luz (LED) blanca, recubierto con papel celofán amarillo, para atenuar el intenso componente azul. El LED se alimentó con una fuente regulada de 5 voltios, con un potenciómetro en serie para controlar la corriente suministrada, entre 1 y 20 mA (Rossi et al., 2013).

La figura 11 es una fotografía del espectro de la luz transmitida cuando el porta-muestras tiene una cubeta descartable conteniendo agua (blanco) (Rossi et al., 2013).

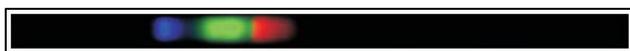

**Figura 11. Espectro visible un LED de luz blanca, cubierto por celofán amarillo**

En la figura 11, cada columna de píxeles contiene información cuantitativa de la intensidad de luz registrada en función de la longitud de onda. Existen sin embargo algunos detalles que considerar (Rossi et al., 2013).

Aunque esté impresa en escala de grises, se trata de una fotografía a color y está constituida por tres canales Rojo (R), verde (G) y Azul (B), que para trabajo cuantitativo es conveniente considerarlas como tres fotografías independientes como se muestra en la figura 12.

En el estándar RGB, la relación entre la potencia de radiación $P_i$, recibida por un píxel dado y el valor numérico $N_i$ registrado (para un canal dado) no es lineal, sino que obedece a la relación:
$N = (P)^{0,45}$ (Rossi et al., 2013).

En el estándar RGB de 24 bits, los valores permitidos para $N_i$ corresponden al intervalo [0, 255], para cada canal (Rossi et al., 2013).

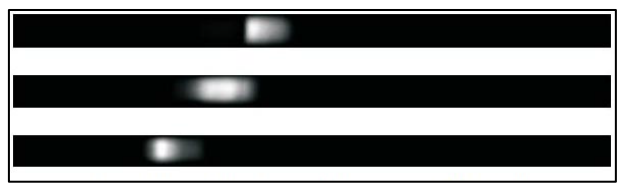

**Figura 12. Canales R, G y B, correspondientes a la imagen 10.**

Estas limitaciones de una cámara web, además de una relación señal / ruido algo desfavorable para trabajo fotométrico, condujeron al desarrollo del siguiente procedimiento para el procesamiento de imágenes:

- Adquirir un conjunto de 128 fotografías para el blanco.
- Separar cada fotografía en sus respectivos canales R, G y B.
- Convertir cada fotografía R, G y B a formato de 32 bits.
- Elevar todos los píxeles de cada fotografía a la potencia de $(1/0,45) = 2,2222$, como se indica en Suzuki et al10.
- Para cada canal, sumar las 128 fotografías (píxel a píxel) y obtener una imagen promedio.
- Sumar las imágenes promedio, correspondientes a los canales R, G y B, para obtener una imagen total, en escala de grises, con profundidad de 32 bits.
- Repetir los pasos 1 a 6, para la muestra y el fondo (cubeta conteniendo el blanco y fuente de luz apagada).
- Obtener la imagen neta del espectro del blanco, restando la imagen promedio del fondo de la imagen promedio del blanco.
- Obtener la imagen neta del espectro de la muestra, restando la imagen promedio del fondo de la imagen promedio de la muestra.
- Poner el valor mínimo de cada imagen neta en 1, para evitar errores de procesamiento, causados por división entre cero y por logaritmo de cero.
- Dividir, píxel a píxel, la imagen neta del espectro del blanco entre la imagen neta del espectro de la muestra. El resultado es la imagen neta de absorción.
- Tomar el logaritmo, base 10, de la imagen neta de absorción. El resultado es la imagen neta de absorbancia.

En el caso de espectrometrías de emisión, se siguen los pasos 1 al 9, omitiendo el paso 8 si es que no se emplea un blanco. Para obtener una curva de calibración fotométrica se sigue los mismos procedimientos, pero empleando soluciones patrón en lugar de muestra (Rossi et al., 2013).

La calibración del espectrómetro en longitudes de onda se realizó empleando como patrón el espectro discontinuo de emisión, de una lámpara fluorescente compacta.

Para la obtención de espectros de absorción se usó soluciones diluidas de permanganato de potasio de diferentes concentraciones, que se obtuvieron mediante diluciones, a partir de una solución madre, de concentración arbitraria.

Para los experimentos de fluorescencia molecular se obtuvo una solución fluorescente, de concentración arbitraria, sumergiendo en agua destilada la punta de un plumón marcador amarillo. Finalmente, para la obtención del espectro de emisión atómica del sodio, se introdujo una pequeña gota de solución diluida de cloruro de sodio, en la llama de un mechero de alcohol, empleando un alambre de nicromo.

*4.6. Espectrofotómetro de absorción cuantitativa de bajo costo*

Los espectrofotómetros modernos de menor costo (por ejemplo, Vernier SpectroVis Plus) cuestan aproximadamente $557 dólares americanos más impuestos de importación y también requieren una PC para la adquisición y manipulación de datos (Albert et al., 2012).

Para ayudar a abordar esta deficiencia, se desarrolló un Sistema simple, robusto y de bajo costo hecho en casa espectrofotómetro visible.

El diseño utiliza una batería, un diodo emisor de luz blanca (LED) con una resistencia limitadora de corriente (similar a las de LED cenizas), una lente, una cubeta y portacubetas, una rejilla montada en un portaobjetos, un fotodiodo montado en un brazo giratorio mediante una simple bisagra de acero y un multímetro digital. Para facilitar la alineación y el uso, hicimos un " mesa óptica " utilizando la mitad de una placa base LEGO de 10 x 10 pulgadas y bloques de construcción LEGO (Albert et al., 2012).

El enfoque principal para el desarrollo del instrumento fue la simplicidad, la facilidad de construcción y uso, y la minimización de costos, sin sacrificar rendimiento. Todo el aparato se construyó sobre un tubo de 5 pulg. × placa base LEGO de 10 pulg. con bloques LEGO (Albert et al., 2012).

La fuente de luz era un LED blanco de 5 mm de diámetro y 50 $\Omega$ resistencia limitadora de corriente, alimentada por tres pilas AA. Aunque hay una lente integrada en el paquete LED, se agregó una segunda lente para colimar mejor la luz. La luz blanca atraviesa la muestra, que se mantiene en un plástico de 1 cm. × Cubeta de 1 cm. La luz transmitida se dispersa mediante un dispositivo montado en diapositivas de 1000 líneas / mm. de la rejilla de difracción. La luz refractada se detectó con un fotodiodo montado en un brazo giratorio. Las longitudes de onda se calcularon directamente desde el ángulo del detector utilizando la ecuación de refracción. El voltaje producido por el fotodiodo se leyó usando un multímetro digital (Albert et al., 2012).

## 5. Aplicaciones

*5.1. Salud general*

Los espectrofotómetros se han utilizado en la atención médica ya que permite realizar algunos tipos de análisis no invasivos, simples y de bajo costo. Pueden ser utilizados para monitorear el medio ambiente en busca de alguna fuente de riesgo para la salud al inhalar algún material particulado (PM) (Kawamura et al., 2006), pesticida (Kawamura et al., 2006) o algún compuesto orgánico volátil (COV) (Smiddy, Papkovskaia, Papkovsky, & Kerry, 2002).

Existe un estudio que es el síndrome del edificio enfermo (SBS), es un problema de salud común en ciudades grandes. Esta situación se presenta cuando los ocupantes de un edificio experimentan algún efecto en su salud que se relacione con el tiempo que han vivido en el complejo, en ocasiones no se puede detectar la sustancia que puede causar los problemas en la salud (Smiddy et al., 2002).

El tolueno es un contaminante que se encuentra asociado a causar el Síndrome del Edificio Enfermo SBS por sus siglas en ingés, es una situación en la que los ocupantes experimentan efectos agudos que pudieran estar asociados con el tiempo de permanencia en el edificio (García, Arriaga, Alatriste, & Aizpuru, 2012), para poder detectar este compuesto Kawamura desarrolló un sensor pudo detectar gas tolueno en el aire a una concentración de 0.05 ppm, que estaba por debajo del valor de concentración de referencia de 0.07 ppm estipulado por la OMS.

La industria alimenticia también se encuentra ligada a la salud ya que las enfermedades como como la fiebre aftosa, el virus de la gripe aviar y la encefalopatía espongiforme bovina, han provocado que exista una garantía de que los alimentos sean saludables y para ello, los sensores químicos basados en LED han sido utilizados para ello(Firstenberg-Eden & Shelef, 2000; Pacquit et al., 2007, 2006). Un ejemplo consiste en que Pacquit empleó un detector basado en LED para monitorear el deterioro del pescado (Ichikawa, Horiuchi, Ichiba, & Matsumoto, 1990; Pacquit et al., 2006).

En el campo de la medicina se han empleado durante mucho tiempo en el diagnóstico médico no invasivo mediante sensores fotodetectores (Mitrani et al., 1991; Teshima, Li, Toda, & Dasgupta, 2005). Un ejemplo de esto es que en 1991 Mitrani investigó el uso de un detector LED para realizar reografía por reflexión de la luz (LRR), que permitía el diagnóstico de trombosis venosa profunda (TVP) (Teshima et al., 2005).

Como se puede observar la espectroscopia en el campo de la salud abarca gran cantidad de ramas que van desde la medicina hasta el medio ambiente, lo cual hace que el continuo desarrollo de espectrofotómetros portátiles y económicos contribuyan en el área de la salud.

*5.2. Medio ambiente*

Para el control ambiental los espectrofotómetros son un herramienta de gran utilidad y más aún si se pueden hacer portátiles como en el caso de Worsfold que ha empleado con éxito sensores químicos basados LED para el seguimiento de una variedad de analitos con un enfoque particular en fosfato (Blundell, Worsfold, Casey, & Smith, 1995; Ellis, Lyddy-Meaney, Worsfold, & McKelvie, 2003; Gardolinski et al., 2002; Trojanowicz, Worsfold, & Clinch, 1988; P. J. Worsfold, Richard Clinch, & Casey, 1987)y nitrato / nitrito / amoniaco (Andrew, Worsfold, & Comber, 1995; Benson, McKelvie, Hart, Truong, & Hamilton, 1996; Clinch, Worsfold, & Casey, 1987; David, McCormack, Morris, & Worsfold, 1998).

O'Toole aplicó un diodo emisor-detector emparejado (PEDD) a la detección de fosfato utilizando el método espectrofotométrico del verde de malaquita (O'Toole, Lau, Shepherd, Slater, & Diamond, 2007).

Como se observó con los espectrofotómetros analizados en este artículo se pueden realizar mediciones químicas del agua que permiten saber si esta se encuentra contaminada o no, demostrando la utilidad de los espectrofotómetros en la rama del ambiente.

*5.3. Seguridad*

En cuestión de seguridad la detección de explosivos y materiales ilícitos relacionados con explosivos es un área importante el monitoreo continuo de explosivos, ya que ofrece una advertencia rápida. La identificación y cuantificación de explosivos es de suma importancia en la práctica forense, las actividades antiterroristas y la detección de minas (Lee, Aldis, Garrett, & Lai, 1982; Pamula, Srinivasan, Chakrapani, Fair, & Toone, 2005).

Lu y col. desarrollo un microchip de electroforesis capilar (CE) para la separación y detección colorimétrica de tres explosivos trinitroaromáticos en agua de mar. El sensor empleado para la detección de los explosivos trinitroaromáticos consistió en un LED verde (λ max505 nm) como fuente de luz y un tubo fotomultiplicador desplazado al rojo en miniatura como detector. Al acoplar la separación de microchip con extracción en fase sólida (SPE), se lograron LOD de 0,34, 0,25 y 0,19 µg L -1 para TNT, TNB y tetrilo, respectivamente (Lu, Collins, Smith, & Wang, 2002).

Los agentes de guerra biológica requieren de una detección e identificación de virus demasiado rápida (Higgins et al., 2003). Se han podido desarrollar espectrofotómetros portátiles de alta sensibilidad, como el "Analizador de Ácido Nucleico Avanzado de Mano" (HANAA), que utilizó dos LED, detectó con éxito ADN extraído de colonias de B. anthracis (Lyddy-Meaney, Ellis, Worsfold, Butler, & McKelvie, 2002).

## 6. Conclusiones

En este artículo se abordó la importancia, aplicaciones e implementación de un espectrofotómetro. Entre las aplicaciones más destacadas esta la determinación de la concentración de un compuesto en solución que se basa en absorción de las radiaciones electromagnética, identificación de compuestos por su espectro de absorción. Se analizó los prototipos de espectrofotómetros que se han construido en diversas instituciones, se estudiaron la calidad-precio de los componentes para la construcción de un espectrofotómetro UV-vis. El empleo de luz LED es la fuente de iluminación con las mejores características ya que no consumen mucha energía, son pequeños y pueden transmitir diferentes longitudes de onda. Se propone utilizar un LED UV 355 nm de longitud de onda como en el artículo (Kvittingen et al., 2016) debido a su precio y bajo riesgo para la salud.

Para el rango visible lo más practico es utilizar un LED de luz blanca y un monocromador económico, como puede ser la parte de debajo de un disco y un pequeño motor para seleccionar la longitud de onda como se realizó en el artículo (D. González-Morales et al., 2020a). Para poder medir la transmitancia, sería necesario utilizar dos sensores uno para el espectro visible que sería el sensor de luz ambiental TEMT6000 (D. González-Morales et al., 2020a) y otro para medir la luz UV que sería el sensor de luz UV Gy-8511 Ml8511 (Mercado Libre, 2020) ya que no son demasiado costosos y funcionan correctamente.

El NodeMCU (Chaianantakul et al., 2018) será utilizado para para realizar los cálculos y que los resultados puedan ser enviados a un servidor web, para poder almacenar la información de forma práctica y remota.

Para la fuente de alimentación se emplearan pilas de 9V o un cargador de celular adaptado (Kvittingen et al., 2016) que permiten la alimentación del NodeMCU, el cargador permite evitar variaciones en los resultados ocasionadas por el desgaste de las pilas.

Por último para un ensamblaje profesional de los componentes debe realizar los planos e imprimirlos en una impresora 3D (Pereira & Hosker, 2019), a falta de impresora 3D se pueden usar bloques LEGO (Pereira & Hosker, 2019), (Kvittingen et al., 2016).